\documentclass[a4paper,fleqn,useAMS,usenatbib]{mnras}

\pdfoutput=1

\usepackage{natbib}
\usepackage{mathptmx}
\usepackage{verbatim}

\usepackage[T1]{fontenc}
\usepackage{ae,aecompl}

\usepackage[dvips]{graphicx}
\usepackage{amsmath}
\usepackage{amsfonts}
\usepackage{amssymb}
\usepackage{comment}
\usepackage{bm} 
\usepackage[dvipsnames]{xcolor}
\usepackage[%
        ]{hyperref}
        
\hypersetup{
	colorlinks=true,	
	urlcolor=MidnightBlue,		
	pdfpagelabels=true,
	hypertexnames=true,
	plainpages=false,
	naturalnames=true,
	pdftitle={The multiplicity distribution of Kepler's exoplanets},    
	pdfauthor={Kipping \& Sandford},     
	linkcolor=WildStrawberry,          
	citecolor=ForestGreen,        
}

\newcommand{\kepler}{{\it Kepler}}
\newcommand{\forecaster}{{\tt forecaster}}
\newcommand{\keplerports}{{\tt KeplerPORTs}}

\newcommand{\pdf}{\mathrm{Pr}}

\title[The \kepler\ multiplicity distribution]{
The multiplicity distribution of \kepler's exoplanets
}
\author[Sandford et al.]{Emily Sandford$^{1}$\thanks{E-mail:
\href{mailto:esandford@astro.columbia.edu}{esandford@astro.columbia.edu}}, David Kipping$^{1}$ and Michael Collins$^{2}$\\
$^{1}$Dept. of Astronomy, Columbia University, 550 W 120th Street, New York NY 10027\\
$^{2}$Dept. of Computer Science, Columbia University, 1214 Amsterdam Avenue, New York NY 10027}

\date{Accepted . Received ; in original form }

\pubyear{2019}

\begin{document}
\label{firstpage}
\pagerange{\pageref{firstpage}--\pageref{lastpage}}
\maketitle

\begin{abstract}
The true multiplicity distribution of transiting planet systems is obscured
by strong observational biases, leading low-multiplicity systems to be
overrepresented in the observed sample. Using the \kepler\ FGK planet hosts,
we employ approximate Bayesian computation to infer the multiplicity
distribution by comparing simulated catalogs to the observed one. After
comparing a total of ten different multiplicity distributions, half of which
were two-population models, to the observed data, we find that a single-population
model following a Zipfian distribution is able to explain the \kepler\ data as
well as any of the dichotomous models we test. Our work provides another example of 
a way to explain the observed \kepler\ multiplicities without invoking a dichotomous 
planet population. Using our preferred Zipfian
model, we estimate that an additional $2393_{-717}^{+904}$ planets likely reside
in the 1537 FGK \kepler\ systems studied in this work, which would increase the
planet count by a factor of $2.22_{-0.36}^{+0.46}$. Of these hidden worlds,
$663_{-151}^{+158}$ are expected to reside in ostensibly single-transiting-planet
systems, meaning that an additional planet(s) is expected for approximately
1-in-2 such \kepler\ systems.
\end{abstract}

\begin{keywords}
planets and satellites: dynamical evolution and stability --- methods: numerical --- stars: planetary systems
\end{keywords}

\section{Introduction}
\label{sec:intro}

From our vantage point within the Solar System, it seems natural to expect
that stars should be accompanied by multiple planets. Around stars similar
to the Sun, planetary systems are common, with numerous studies of the
\kepler\ sample converging on an occurrence rate of at minimum one planet
per star \citep{petigura:2013,dfm:2014,burke:2015,hsu:2018}.
In the majority of \kepler\ systems, there is just a single transiting
planet detection \citep{thompson:2018} but the strong observational biases
plaguing transit surveys \citep{sandford:2016} mean that one might reasonably
expect many of these to in fact be multi-planet systems (``multis'') yet
to be revealed.

Measuring the multiplicity distribution is crucial, as it is a vital
clue to the origins and evolution of detected systems. For example,
hot Jupiters rarely reside in multi-planet systems \citep{wright:2009,
steffen:2005,gibson:2009,latham:2011,steffen:2012} barring exceptional
cases like WASP-47b \citep{becker:2015,weiss:2017}. This is often
interpreted as evidence for late inward migration from beyond the snow line,
leading to scattering of interior planetesimals \citep{beague:2012,
spalding:2017,heller:2018,dawson:2018}. At the other extreme, systems like
Kepler-11 \citep{lissauer:2011a} and TRAPPIST-1 \citep{gillon:2017} pack
half a dozen planets within the orbit of Mercury, which suggests that disk
migration and resonant trapping may guide the evolution of such systems
\citep{quillen:2006,mustill:2011,ormel:2017,tamayo:2017,papaloizou:2018}.

Unfortunately, the multiplicity distribution is not directly observable from
transit surveys like \kepler, because of the extreme biases inherent to the
technique. Nevertheless, the unparalleled volume and homogeneous detection
biases of the \kepler\ planets still make it arguably the best resource for the
task. If we imagine a system of multiple planets around a star, it is likely that
only a subset of the planets (if any) will be detected by a transit
survey such as \kepler, because it is guaranteed neither that all of the planets
will be aligned with our line of sight, nor that all will transit at high enough
signal-to-noise to be detected. Furthermore, \cite{zink19} investigate the detection efficiency of the \textit{Kepler} pipeline and find that it drops in multi-planet systems---specifically, the detection efficiency is higher for the first transiting planet discovered around a star than for subsequent transiting planets in the same system.

Unveiling the true multiplicity distribution from
the observed one is therefore a challenging task that needs to account for
both geometric and detection biases. Specifically, the mutual
inclination distribution between planets should be taken into account in any
exploration, since it combines with the underlying multiplicity distribution to 
produce the observed catalog \citep{tremaine:2012,brakensiek:2016}.

One of the first attempts to model the \kepler\ multiplicity distribution is presented
by \citet{lissauer:2011b}, who test a Poisson multiplicity model and
find that the \kepler\ catalog is best fit when the mean of this model is equal to 5.5
planets, and the mutual inclinations of the planets are low. However, they find that
this best-fitting model significantly underpredicts the number of
single-planet systems observed by \kepler. The case for 
low mutual inclinations in particular has
been reproduced in numerous studies \citep{fang:2012,figueira:2012,
weissbein:2012,fabrycky:2014}. \citet{ballard:2016}, who study the \kepler\ 
M-dwarfs, suggest that the
under-prediction of single-planet systems can be resolved by introducing a 
dichotomous population, with one component being a dynamically cold set of
multis and the second being a population of singles or highly-mutually-inclined multis.

This \kepler\ dichotomy, if it extended to FGK stars, may also explain the 
under-prediction of singles observed by \citet{lissauer:2011b}. This has motivated
follow-up efforts to determine if there are fundamental differences in the
stellar hosts between singles and multi-planet systems. In particular,
\citet{romero:2018} search for, but ultimately find no evidence of,
metallicity differences among the hosts of the two types of systems, which might be expected if giant planets were responsible for the dynamically hot population.

The lack of any significant metallicity difference has led some to question
the dichotomous hypothesis, even as it explains the observed M-dwarf multiplicities better 
than a single-population model. The drop in detection efficiency for subsequent planets detected in multi-planet systems noted by \cite{zink19} hints at another explanation for the overabundance of observed singles, that the \textit{Kepler} pipeline simply fails to detect subsequent planets around some percentage of ``singles." Indeed, after accounting for this effect, \cite{zink19} find that a modified Poisson model fits the observed multiplicities of \textit{Kepler} GK stars well.

Another alternative explanation to a dichotomous model is a flatter inclination distribution in the inner parts of multi-planet systems than previously assumed---because this inclination distribution works together with the underlying multiplicity distribution to create the observed multiplicities, it is important to model both \citep{tremaine:2012}. 
\citet{bovaird:2017} show that by adopting a flat-disk model, rather than a ``flared" disk, they can match the observed \textit{Kepler} multiplicities without invoking a dichotomous model. 

In this work, we aim
to address the question of multiplicity by presenting a comparison of
several plausible multiplicity distribution models, including both single
and dichotomous populations.

We structure this paper as follows. In Section~\ref{sec:methods}, we
introduce the methods, models and inference approach of this work.
In Section~\ref{sec:analysis}, we present the results, visualizations
and analysis of the fits. Finally, we place our work in a broader
context in Section~\ref{sec:discussion}.

\section{Methods}
\label{sec:methods}

\subsection{Input catalog}
\label{sub:catalog}

We downloaded the \kepler\ DR25 Kepler Objects of Interest (KOIs) catalog via the
NASA Exoplanet Archive (NEA; \citealt{akeson:2013}), with several
filters applied. First, we selected only KOIs for which the ``Disposition
using Kepler Data'' was reported as ``CANDIDATE''. Second, we required that the
NEA-reported surface gravity of the star satisfied $\log g > 4$ and that the
stellar mass was $0.8 < (M_{\star}/M_{\odot}) < 1.2$, in order to focus on FGK
dwarfs. Finally, we filtered for planetary candidates which satisfied
$6.25<(P/\mathrm{days})<400$ (the same range considered by
\citealt{petigura:2013} and \citealt{dfm:2014}) and $0.5<(R_P/R_{\oplus})<32$.
This led to a population of 1966 KOIs, of which the majority were dispositioned
as ``CONFIRMED''.

These 1966 KOIs define our observed data set, $\mathcal{D}_{\mathrm{obs}}$,
 which comprises three key pieces of information. First,
the observed multiplicity distribution, which is simply the occurrence tally
of multiplicities from 1 to 10 and is reported in Table~\ref{tab:kepcounts}.
Second, the list of maximum a-posteriori probability orbital periods, of which there are
1966 elements. Third, the list of maximum a-posteriori probability planetary radii,
of which again we have 1966 entries.

\begin{table}
\caption{\emph{Observed multiplicities in the final subset of 1966 KOIs considered in this work.
Taking the sum of each multiplicity by its count yields 1966, as expected.}} 
\centering 
\begin{tabular}{c c c c c} 
\hline\hline 
Multiplicity, $m$ & Counts, $n_{\mathrm{obs},m}$ \\ [0.5ex] 
\hline 
1 & 1225 \\
2 & 218 \\
3 & 76 \\
4 & 15 \\
5 & 1 \\
6 & 2 \\
7 & 0 \\
8 & 0 \\
9 & 0 \\
10 & 0 \\ [1ex]
\hline\hline 
\end{tabular}
\label{tab:kepcounts} 
\end{table}

After compiling $\mathcal{D}_{\mathrm{obs}}$, we also queried all \kepler\
target stars for stars which match the filters imposed above.
For this, we took the \citet{mathur:2017} DR25 catalog of stellar parameters,
which listed 197,096 stars, and cross-matched these with the CDPP$_6$ values
(combined differential photometry on a 6-hour timescale; see
\citealt{christiansen:2012}) as obtained from the Mikulski Archive for Space
Telescopes (MAST). The mean CDPP$_6$ across all quarters of a given star was
saved as the representative CDPP$_6$. In some rare instances, these values were
not available on MAST and thus these stars were dropped, leaving us with
196,792 stars. We then applied the same cuts for $\log g$ and $M_{\star}$ as
described in the previous paragraph, leaving us with 108,429 FGK dwarfs. This
catalog of stars will be used later in Section~\ref{sub:forward}.

\subsection{Tackling completeness}
\label{sub:completeness}

Our objective is to infer the multiplicity distribution from the \kepler\
catalog. At a very basic level, this objective is challenged by the
incompleteness of the \kepler\ catalog itself - just because a
star has a planet doesn't mean \kepler\ is guaranteed (or even likely) to
see it. A great deal of attention has been paid to this issue in connection
to estimating the underlying planet occurrence rate from \kepler, and so
although our objective is distinct, it is useful to briefly review the
approaches used in such studies a source of guidance.

The simplest form of incompleteness to deal with is the geometric transit
probability, which decreases with increasing planet orbital radius. In
estimates of \kepler\ planet occurrence rates, this can be
most easily accounted for by simply dividing apparent occurrence rates by
$R_{\star}/a$ (the geometric transit probability), under the assumption of
close-to-circular orbits (e.g. see \citealt{howard:2012}).

The second, and more challenging, component to completeness is detection
efficiency. The simplest solution is to limit one's analysis to a parameter
subset where one assumes that the completeness is approximately unity (e.g.
\citealt{howard:2012,fang:2012}). This naturally comes at the expense of
a smaller sample size. In order to expand the sample to lower signal-to-noise
ratio (SNR) events, it is necessary to estimate the detection efficiency in
more detail. A typical approach is the so-called ``inverse detection efficiency
method'' (IDEM)\footnote{As dubbed by \citet{dfm:2014}.}, where each planet
is assigned a detection efficiency score and ultimately the true occurrence
rate is inferred by dividing by both the transit probability and the detection
efficiency. Detection efficiencies are typically estimated by injection and
recovery exercises (e.g. see \citealt{petigura:2013,dressing:2015}). As an
example, for a given choice of orbital period and planetary radius, the
associated detection efficiency for a particular star may be computed using
the \keplerports\ software \citep{burke:2015,burke:2017}.

In the case of planet occurrence rate estimation, the simplest strategy to
account for detection efficiency is the IDEM approach. However, recently
\citet{hsu:2018} argue that this approach leads to systemic biases in the
inferences since the efficiencies are drawn from estimated planet properties,
which are themselves uncertain. Instead, they use a forward-model to inject a
population, filter it through a realistic detection efficiency model, and then
compare the surviving population to the observed population with some distance
metric and Approximate Bayesian Computation (ABC). This approach is shown to
more faithfully infer the occurrence rate of the injected population and so a
similar approach is adopted here. Rather than comparing occurrences, our
work ultimately is interested in the frequency of various multiplicities, but
the same approach can be employed (and is discussed further in
Section~\ref{sub:regression}).

In principle, it should be possible to define a detection efficiency model
unique to each star using \keplerports. However, when conducting Bayesian
inference, detailed calculations of these efficiencies for every star and at
every step in period and radius comes at high computational cost. A simpler yet
still accurate approach is to use a global \kepler\ detection efficiency model,
for which one inputs the so-called multiple event statistic (MES) of a
planetary candidate and the model returns a detection probability. This is
appropriate since we primarily care about the ensemble rather than individual
systems. One example of a global detection efficiency model comes from
\citet{christiansen:2016}, who use the actual \kepler\ detection pipeline to
inject and recover and planets and find that, for FGK stars, the average
detection efficiency is well-approximated by either a cumulative gamma function
or a logistic function. We use the latter in this work since it is faster to
compute. It is given by

\begin{align}
\mathrm{Pr}(\mathrm{detection}|\mathrm{MES}) &= d_l - \frac{d_l}{1+(\mathrm{MES}/c_l)^{b_l}},
\label{eqn:completeness}
\end{align}

where $b_l$, $c_l$ and $d_l$ are coefficients defined in
\citet{christiansen:2016} using \kepler\ DR24. An update to the cumulative
detection fraction versus MES is presented in \citet{thompson:2018} for DR25,
who find a similar distribution which we use in this work.

The MES (see \citealt{jenkins:2002}) is a statistic that measures the combined
significance of all the observed transits in the detrended, whitened \kepler\
light curves, assuming a linear ephemeris. In practice, it is not feasible
to generate very large populations of synthetic planets (required for Bayesian
Monte Carlo work), inject their transits, detrend, whiten, and thus compute MES
in the same way as the real \kepler\ pipeline.

Instead, we use the transit SNR as a proxy for the MES in what follows.
We note that the two are not equal---specifically, the MES depends on the
goodness-of-fit between a transit search template and a transit signal mediated
by stellar noise, while the SNR does not depend on the template---but they are,
to first order, proportional to each other \citep{burke:2017}. By combining the
SNR with Equation~(\ref{eqn:completeness}), we are able to estimate the
detectability of any synthetic KOI.

\subsection{The forward model}
\label{sub:forward}

Our model works by first choosing a random star from the filtered stellar
catalog described in Section~\ref{sub:catalog}. We then inject a planetary
system around it composed of $m$ planets, where $m$ is always less than or
equal to $m_{\mathrm{max}}=10$ and is drawn from a chosen multiplicity
distribution as described later in Section~\ref{sub:models}. Each of the
planets is assigned a random period drawn from a log-uniform distribution from
6.25 to 400\,days. A log-uniform distribution was chosen since it both provides
a reasonably close match to the observed marginalized period distribution
reported by \citet{dfm:2014}, and is the same assumption used in previous
multiplicity studies, such as \citet{ballard:2016}. 

Next, the innermost planet in the system is assigned a random radius drawn from a double-sided power
law (DSPL) distribution, described by

\begin{equation}
\mathrm{Pr}(R) \propto
\begin{cases}
(\log R - \log R_{\mathrm{min}})^{-\alpha_{\mathrm{small}}}  & \text{if } R_{\mathrm{min}}<R \leq R_{\mathrm{crit}} ,\\
(\log R - \log R_{\mathrm{crit}})^{-\alpha_{\mathrm{big}}}  & \text{if } R_{\mathrm{crit}}<R<R_{\mathrm{max}} .\\
\end{cases}
\end{equation}

We normalize the DSPL distribution such that the two sides meet at
$R_{\mathrm{crit}}$ and integrate to unity over the interval
$R_{\mathrm{min}}<R<R_{\mathrm{max}}$. The terms $R_{\mathrm{min}}$ and
$R_{\mathrm{max}}$ are fixed to 0.5\,$R_{\oplus}$ and 32\,$R_{\oplus}$
respectively, but the parameters $R_{\mathrm{crit}}$, $\alpha_{\mathrm{small}}$
and $\alpha_{\mathrm{big}}$ are treated as unknown shape parameters to be
inferred.

To reflect the observed ``peas-in-a-pod" covariance of planet radii
\citep{weiss:2018}, the radii of subsequent planets in the system are drawn
from a Gaussian distribution centered at the innermost planet's radius. The
scale parameter of this distribution, $\sigma_R$, is treated as another free
parameter to be inferred. This parameter is able to extend out to very large
values, thereby accounting for the possibility of no correlation
\citep{zhu:2019}.

The simplified double-sided power law radius distribution is designed to
capture the turn-over in planet occurrence seen at around mini-Neptune radii,
reported in numerous studies \citep{fressin:2013,petigura:2013,dfm:2014}. It
does not, however, describe the radius valley reported by \citet{fulton:2017}.
This effect was only revealed by substantial improvements to the precision of
measured stellar radii, and we argue that it is not influential enough to
significantly affect our study which focusses on multiplicity. 

The radius distribution used here essentially represents a set of nuisance
parameters which is marginalized over in the final results.


Having generated $m$ proposal planets around the star, with periods and radii
drawn from the distributions described above, we next check whether the system
is dynamically stable. Following the same approach as \citet{ballard:2016}, we
test for Hill stability using Equation~(3) of \citet{fabrycky:2014}.
Specifically, we define the mutual Hill radius between planets ``1'' and ``2''
as

\begin{align}
R_H &= \Bigg(\frac{M_1 + M_2}{3 M_{\star}}\Bigg)^{1/3} \frac{ a_1 + a_2 }{2},
\end{align}

where the Hill stability criterion is satisfied if

\begin{align}
\frac{ a_2 - a_1 }{ R_H } > \Delta_{\mathrm{crit}},
\end{align}

where $M$ and $a$ refer to the masses and semi-major axes of the planets. To
estimate $M$ for each planet, we use the maximum
a-posteriori probability \forecaster\ mass-radius relation derived by \citet{chen:2017}.
We compute semi-major axes are from periods using the stellar mass and Kepler's
Third Law.

The critical separation is $\Delta_{\mathrm{crit}}=2\sqrt{3}$ for neighboring
planets, and for three-or-more planets, \citet{fabrycky:2014} require
$\Delta_{\mathrm{inner}} + \Delta_{\mathrm{outer}} > 18$ for neighboring inner
and outer pairs of planets.

If the proposed planetary system violates Hill stability, we use the same star
and same multiplicity but make a new realization of the periods and radii for
the planetary system. We allow this process to repeat up to 1000 times, after
which we abandon the star and draw a new star from the KIC catalog.

After this point we have generated a stable multi-planet system. Next, we
need to calculate how many of these planets actually transit. The innermost
planet, labelled with subscript ``1'', has a transit impact parameter, $b_1$,
of $(a_1/R_{\star}) \cos I_1$, where $I_1$ is the orbital inclination
angle\footnote{Note that a near-circular orbit is assumed here and
throughout.}. Inclination is isotropically distributed and thus we adopt a
uniform distribution for $\cos I_1$, which in practice means that we draw a
random real number for $\cos I_1$ from $\mathcal{U}[0,1]$, where $\mathcal{U}$
denotes a uniform distribution. The other planets in the system are assumed to
have inclinations perturbed away from this angle by an angle $\Delta I$, representing their
mutual inclinations within a flared disk (see \citealt{bovaird:2017} for a
flat-disk model). $\Delta I$ is drawn from a Rayleigh distribution
characterized by a scale parameter $\sigma_I$:

\begin{align}
\mathrm{Pr}(\Delta I) &= \frac{\Delta I}{\sigma_I^2} \exp\Bigg(\frac{-\Delta I^2}{2\sigma_I^2} \Bigg)
\end{align}


For each planet in the system, we calculate  the impact parameter
$b_j = (a_j/R_{\star})\cos{( I_1 + \Delta I_j)}$. Any planet for which
$b_j < 1 + (R_j/R_{\star})$ is treated as a transiting planet and is saved.
Systems with zero transiting planets need not be considered further and are
discarded, leading us to draw a new star from the KIC catalog.

At this point, we now have a simulated system of at least one transiting planet 
orbiting a chosen KIC star. The final component of our forward model is to
simulate what fraction of transiting planets in the system would actually be
detectable. To do this, we first assign each planet a random transit epoch.
Next, we query which quarters that particular KIC star was observed by \kepler\
for, since many stars were not observed in every quarter due to spacecraft 
rotation and loss of CCDs during the mission.

Using our simulated ephemeris for each planet, we can now calculate how many
transits of each planet would have been observed by \kepler. We estimate
the SNR of each planet using Equation~(10) of \citet{sandford:2016}, multiplied
by the square-root of the number of observed transits. Finally, the
detection probability is computed using Equation~(\ref{eqn:completeness}) from
\citet{christiansen:2016}. To decide if the transiting planet is detectable or
not, we make a random Bernoulli draw, with probability equal to this computed
detection probability.

This process culminates in a set of $m$ simulated detected transiting planets 
around a particular star. At this point, we loop back to the beginning of the
forward model and keep going until 1966 detected planets have been generated
(since this represents the size of the observed sample, $\mathcal{D}_{\mathrm{obs}}$,
that we will ultimately compare to). 

The forward model therefore ultimately yields a simulated data
set, $\mathcal{D}_{\mathrm{sim}}$, with the same elements and form as
$\mathcal{D}_{\mathrm{obs}}$. Further,$\mathcal{D}_{\mathrm{sim}}$ is clearly 
dependent upon the simulation's choice of multiplicity, radius and inclination
distributions - which are characterized by model parameters $\boldsymbol{\theta} =
\{ \beta, \alpha_{\mathrm{small}},\alpha_{\mathrm{big}}, R_{\mathrm{crit}}, \sigma_R,
\sigma_I \}$ (where $\beta$ is a stand-in term(s) describing the multiplicity
distribution, described below in Section~\ref{sub:models}, and the other terms have
been previously defined).

We note that since non-detections are discarded and not counted, our approach does
not enable an estimate of the underlying planet occurrence rate.

\subsection{Comparison to observations}
\label{sub:regression}

A single run of the forward model described in Section~\ref{sub:forward}
generates a population of simulated detected transiting planets described by
$\mathcal{D}_{\mathrm{sim}}$. Our task is now to infer the parameters 
of the forward model, $\boldsymbol{\theta}$, which would have generated 
the observed \kepler\ systems, $\mathcal{D}_{\mathrm{obs}}$, by comparing 
$\mathcal{D}_{\mathrm{sim}}$ to $\mathcal{D}_{\mathrm{obs}}$. In particular,
we are interested in inferring the parameters of the multiplicity distribution,
$\beta$.

Conventional Bayesian inference might proceed using hierarchical Bayesian
modeling (HBM), where the multiplicity distribution is described by some
parameterized form and then each system's true multiplicity is treated as a
free parameter drawn from this overall distribution - giving rise to a
large number of unknown variables to solve for (see \citealt{hogg:2010} for
an astronomer's introduction to HBMs). In this case, the
likelihood function used for inference would be well-defined as the product
of the likelihoods for each individual system.

Hierarchical models allow for rigorous inference but typically come at
great computational expense. Instead, we seek to learn the multiplicity
distribution by comparing some distance metric which quantifies how closely the
simulated population matches the observed population - thereby ignoring
the individual systems and treating the population as an ensemble. By using
one or more distance metrics to quantify goodness-of-fit, we are thus
conducting what is typically referred to as approximate Bayesian computation,
or ABC (see \citealt{ishida:2015,hahn:2017,hsu:2018,witzel:2018} for
recent applications in astronomy).

The three key ingredients for ABC inference are a forward model which generates
$\mathcal{D}_{\mathrm{sim}}$, prior distributions for the model parameters, 
$\pdf(\boldsymbol{\theta})$, and a distance function
$\rho(\mathcal{D}_{\mathrm{obs}},\mathcal{D}_{\mathrm{sim}})$ which quantifies
how well the simulated distribution resembles the observed sample.

Although our primary goal is to learn the multiplicity distribution, we elect
to define a distance metric which considers the agreement between the simulated
and observed multiplicities but also the agreement between the simulated and
observed radius distribution. This is because the two cannot be assumed to be
independent: planetary radii determine planetary masses, which in turn determine
their stability and whether they could reside in a high-multiplicity system.
We therefore infer not only the multiplicity model parameters, but also the
radius distribution parameters. Further, the mutual inclination distribution 
strongly influences the fraction of planets observed to transit and thus is also
a parameter we should expect to constrain and be covariant with the other
model terms.

Our goal here, of course, is not to infer the true radius distribution of the
\textit{Kepler} catalog, nor the distribution of mutual inclinations among its
multi-planet systems. However, the known inter-relationships between these terms
necessitates that we have some reasonable description of them and that we freely
explore them in conjunction with the multiplicity distribution. At the end, we
can simply marginalize over these ``nuisance'' terms in our final calculation of
the multiplicity distribution.

Having established that we require a distance metric which incorporates both
the multiplicities and radii, let us consider the multiplicity component first.
Previous works have most commonly invoked a Poisson likelihood function in
comparing a simulated multiplicity to the observed value \citep{weissbein:2012,
ballard:2016}. This essentially asserts that probability distribution for the
observed number of $m$-planet systems, $n_m$, is a Poisson distribution with a
mean rate given by $n_{\mathrm{sim},m}$. The Poisson model is well-motivated
for inference based on counting statistics, such as this, and thus is adopted
in this work too. Accordingly, the probability of observing a particular number
of $m$-planet systems, $n_{\mathrm{obs},m}$, is given by

\begin{align}
\mathrm{Pr}(n_m=n_{\mathrm{obs},m}) &= \frac{ e^{ -n_{\mathrm{sim},m} } n_{\mathrm{sim},m}^{ n_{\mathrm{obs},m} } }{ n_{\mathrm{obs},m}! }.
\end{align}

The Poisson likelihood function is defined by a product of the above over all
$m$ (=multiplicities), and this function certainly describes how close a
simulated set of multiplicities, $\mathbf{n}_{\mathrm{sim}}$, resembles the
observed set, $\mathbf{n}_{\mathrm{obs}}$ - thereby providing a suitable distance
metric:

\begin{align}
\rho(\mathbf{n}_{\mathrm{obs}},\mathbf{n}_{\mathrm{sim}}) &= \prod_{m=1}^{M_{\mathrm{max}}} \frac{ e^{ -n_{\mathrm{sim},m} } n_{\mathrm{sim},m}^{ n_{\mathrm{obs},m} } }{ n_{\mathrm{obs},m}! },
\label{eqn:poissonDistance}
\end{align}

where $\mathbf{n}_{\mathrm{obs}}$ and $\mathbf{n}_{\mathrm{sim}}$ represent
vectors containing all the $m$-indexed observed and simulated population
multiplicities, respectively. The $\mathbf{n}_{\mathrm{obs}}$ vector is fixed 
and given by Table~\ref{tab:kepcounts}. Meanwhile, the
$\mathbf{n}_{\mathrm{sim}}$ vector, which needs to be counted up after each
forward model call, is directly controlled by the choice of forward model
parameters $\boldsymbol{\theta}$,  which we ultimately wish to infer.
It should be noted that our choice of distance metric here, shown in
Equation~(\ref{eqn:poissonDistance}), does not decrease as the distributions
approach one another, but rather increases. Accordingly, it could perhaps be
better thought of as an inverse distance metric although we'll continue to
refer to it as distance metric in what follows, with the only important
consequence being that our task is to maximize $\rho$, rather than minimize
$\rho$.

We now turn our attention to the component of the distance metric which
characterizes the planetary radius distribution. The objective here is not to
fit each and every planetary radius - which would be more in line with an HBM.
Instead, we wish to simulate a population whose statistical properties broadly
match those of the observations. A straightforward approach for accomplishing
this is to the use the Kolmogorov-Smirnov (K-S) test, following on from the
approach adopted by \citet{fang:2012}.

We therefore compute the K-S $p$-value between the observed radii and the
simulated set as our radius distance metric, since this follows the
behaviour of the multiplicity component in terms of being a term we seek to
maximize. We multiply this by the multiplicity distance metric given by
Equation~(\ref{eqn:poissonDistance}) to define an overall distance metric,
$\rho(\mathcal{D}_{\mathrm{obs}},\mathcal{D}_{\mathrm{sim}})$. The two
components are equally weighted under this definition.

A variety of sampling techniques are suitable for ABC inference
\citep{beaumont:2019}, and in this work we elect to use the Markov Chain
Monte Carlo (MCMC) approach \citep{marjoram:2003,marin:2012}.
We sample the model parameter space of $\boldsymbol{\theta}$ with Gaussian
proposals where the acceptance criterion is chosen such that the probability of
accepting a proposal is $\rho_{\mathrm{proposal}}/\rho_i$, where $i$ denotes
the current index in the chain. In this way, improvements in the distance metric
(which recall equates to an increase in the ``distance'' under our definition)
are always accepted. This means that samples near the beginning of the chain,
prior to convergence, can often have poor distance scores. We remove these burn-in
samples by only including samples in the chain past the first instance exceeding
the median distance metric. We demand that 50,000 accepted samples are
achieved for each model, with the final chains inspected to verify
convergence and mixing.

Because of the somewhat subjective nature of choosing an appropriate distance
function in ABC, there is no formal guarantee the model posterior will converge
to the true posterior distribution. For this reason, it is important to test
the accuracy of our ABC inference framework through fake data generation and 
recovery simulations, which we present in Section~\ref{sub:fakes}.

\subsection{Proposed multiplicity models}
\label{sub:models}

In Section~\ref{sub:forward}, we described how exoplanetary radii could be
described using a DSPL distribution, mutual inclinations with a Rayleigh distribution, and orbital periods with a log-uniform distribution.
However, we did not propose a specific form for the multiplicity distribution
itself - which we turn our attention to here.

Specifically, we here describe ten choices of multiplicity distribution: five single-population models, each parameterized by a single free parameter $\beta$, and five corresponding ``dichotomous" models. The dichotomous models are mixture models of two populations: (1) a fraction $f$ of single-planet systems and (2) a fraction $1-f$ of multi-planet systems, distributed according to one of the models parametrized by $\beta$.

\subsubsection{Constant model}
\label{subsub:consant}

One of the first multiplicity models proposed in the literature is presented by
\citet{ballard:2016}, who initially adopt a simple approach where every system
has the same multiplicity, $\beta_{\mathrm{const}}$, which is treated as a free
parameter (we refer to this as the constant model):

\begin{equation}
\pdf(m|\beta_{\mathrm{const}}) =
\begin{cases}
1  & \text{if } m=\beta_{\mathrm{const}} ,\\
0 & \text{otherwise },
\end{cases}
\end{equation}

\citet{ballard:2016} conclude that this model is unable to provide a good
fit to the observed multiplicities of \kepler\ M-dwarf systems and thus
expand upon the constant distribution in that same work to include a second
component of single-planet systems ($m = 1$), which represent a fraction $f$
of all systems:

\begin{equation}
\pdf(m|\beta_{\mathrm{const}},f) =
\begin{cases}
f  & \text{if } m=1 ,\\
1-f  & \text{if } m=\beta_{\mathrm{const}} ,\\
0 & \text{otherwise },
\end{cases}
\end{equation}

Following the terminology used by the authors, we refer to this as the
``dichotomous'' constant model.

Simple models are often attractive since one might plausibly purport that the
laws that govern planetary architectures (or whatever other phenomenon one is
considering) are fundamentally simple themselves. However, we suggest here that
the constant model is almost certainly too simple a model to be a realistic
description of the exoplanet multiplicity distribution. It is rather
implausible to suppose that every system should have an identical number of
planets, and even after including a second population of singletons, this
still yields a highly unnatural distribution composed of two distinct peaks
at $m=1$ and $m=\beta_{\mathrm{const}}$, with zero probability that systems have
multiplicities other than this. For this reason, we felt motivated to consider
other models in addition to the constant formalism.

\subsubsection{Uniform model}
\label{subsub:uniform}

A simple improvement to consider would be to adopt a discrete uniform
distribution, where every multiplicity is just as likely as any other, above
some minimum multiplicity, $\beta_{\mathrm{uniform}}$:

\begin{equation}
\pdf(m|\beta_{\mathrm{uniform}}) =
\begin{cases}
\frac{1}{(m_{\mathrm{max}}+1)-\beta_{\mathrm{uniform}}} & \text{if } \beta_{\mathrm{uniform}} \leq m \leq m_{\mathrm{max}} ,\\
0 & \text{otherwise }.
\end{cases}
\end{equation}

This can be similarly be extended to a dichotomous model by assuming that
some fraction of planetary systems, $f$, belong to a separate population of singles.

\subsubsection{Truncated Poisson model}
\label{subsub:poisson}

Arguably, a more natural model is a Poisson multiplicity distribution,
which appears to be the most commonly adopted law (e.g. see \citealt{lissauer:2011b,fang:2012,gaidos:2016,bovaird:2017}).
This might be expected if the multiplicity 
of an exoplanetary system were the result of a constant rate of generating planets
within a fixed interval of time or space. One may write that the probability of 
forming an $m$-planet system would thus be

\begin{equation}
\pdf(m|\beta_{\mathrm{poisson}}) \propto
\begin{cases}
\frac{ \beta^{ m } }{m!} & \text{if } 1 \leq m \leq m_{\mathrm{max}} ,\\
0 & \text{otherwise },
\end{cases}
\end{equation}

where we drop normalization terms which do not depend on $m$.
For $\beta_{\mathrm{poisson}}\geq2$, this implies a peaked, non-monotonic
distribution at a specific multiplicity, unlike the uniform case. In practice,
we reject any trial $m$ equal to zero or exceeding $m_{\mathrm{max}}$, i.e.
we truncate the distribution. The Poisson can again be extended to a
dichotomous model as was done before. 

\subsubsection{Exponential model}
\label{subsub:exp}

Another previously adopted law is that of a discrete exponential
distribution (e.g. \citealt{bovaird:2017}) which imposes that
the multiplicity, $m$, follows

\begin{equation}
\pdf(m|\beta_{\mathrm{exp}}) \propto
\begin{cases}
\beta_{\mathrm{exp}}^{m} & \text{if } 1 \leq m \leq m_{\mathrm{max}} ,\\
0 & \text{otherwise},
\end{cases}
\end{equation}

which can again be extended to a dichotomous model as above.

\subsubsection{Zipfian model}
\label{subsub:zipf}

Finally, we consider a Zipfian distribution, which represents a discrete
power-law given by

\begin{equation}
\pdf(m|\beta_{\mathrm{zipf}}) \propto
\begin{cases}
m^{-1-\beta_{ \mathrm{zipf} } } & \text{if } 1 \leq m \leq m_{\mathrm{max}} ,\\
0 & \text{otherwise},
\end{cases}
\end{equation}

and a corresponding dichotomous model constructed as above.

Zipf's Law \citep{zipf:1935} is known to be an excellent approximation
for word frequency versus rank for human languages, and even some animal
communications \citep{doyle:2011}. On this basis, it might seem like a
peculiar choice to use when modeling the exoplanet multiplicity distribution,
but Zipf's Law also appears in much wider array of problems, such as
the population frequency of cities \citep{auerbach:1913}. Zipf's Law has
been argued to be a natural by-product of models with many underlying latent
variables, somewhat analogous to the arguments behind the Central Limit
Theorem \citep{belevitch59,aitchison:2016}, and thus on this basis would seem a very
reasonable model to propose for exoplanets too - despite the fact it has
seemingly not been used in the past for this purpose.

\section{Analysis}
\label{sec:analysis}

We run the forward model described in Section~\ref{sub:forward} ten times, once
for each choice of multiplicity model described above, and fit for the free parameters 
$\boldsymbol{\theta} =\{ \beta, \alpha_{\mathrm{small}},\alpha_{\mathrm{big}}, 
R_{\mathrm{crit}}, \sigma_R,\sigma_I \}$ (and $f$, for the dichotomous models)
via ABC, as described in Section~\ref{sub:regression}.

In these fits, we adopt a uniform prior on $R_{\mathrm{crit}}$ between 
$R_{\mathrm{min}} = 0.5 R_{\oplus}$ and $R_{\mathrm{max}} = 32 R_{\oplus}$, and a
uniform prior on $f$ between 0 and 1. We also adopt (improper) priors insisting that
$\alpha_{\mathrm{small}}$ and $\alpha_{\mathrm{big}}$ be greater than $-1$ and 
that $\sigma_R$ and $\sigma_I$ be positive.

\begin{figure*}
\begin{center}
\includegraphics[width=2.05\columnwidth,angle=0,clip=true]{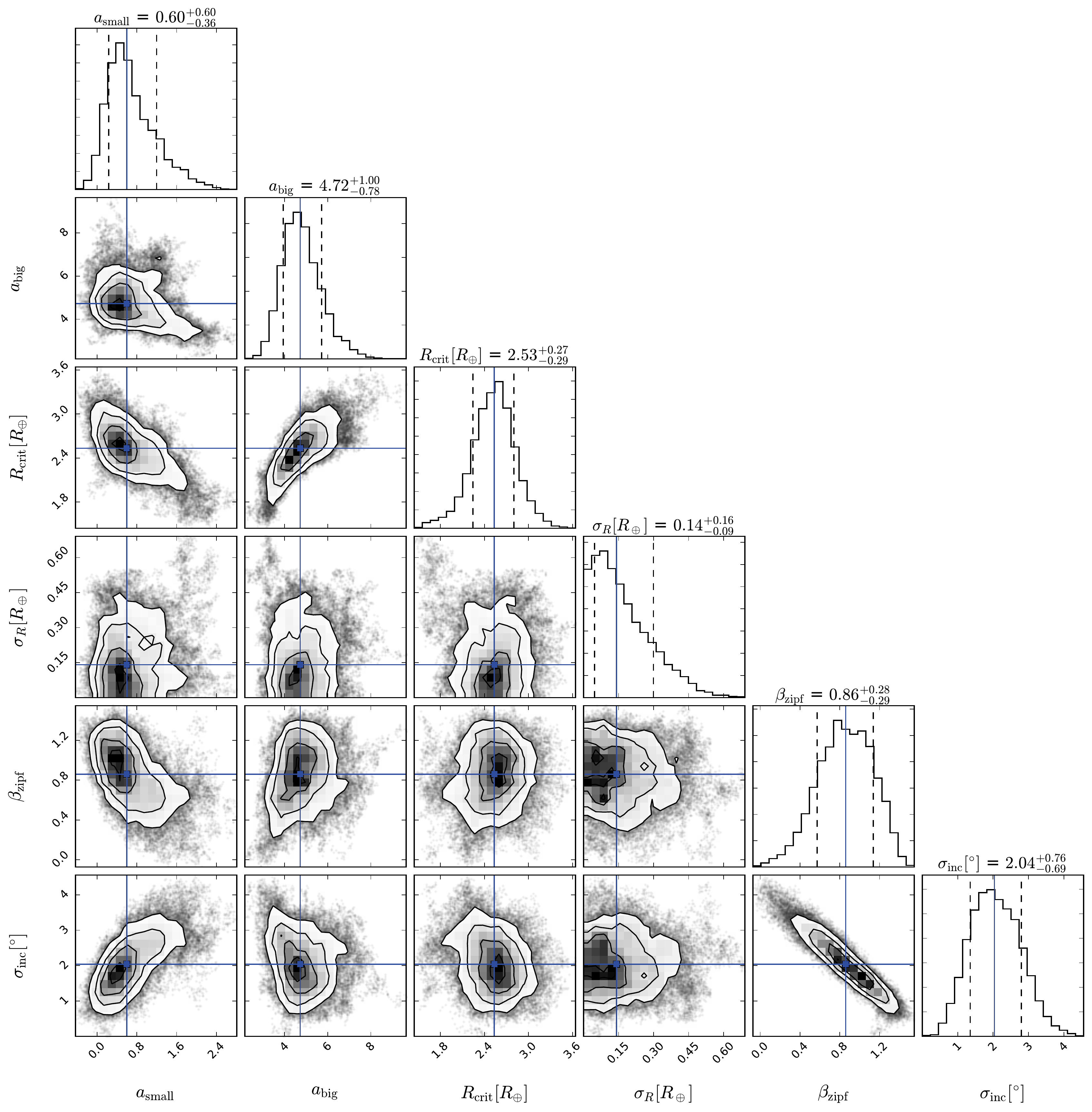}
\caption{\emph{
Joint posterior probability distribution for the non-dichotomous Zipfian
multiplicity model - the favored model deduced in this work. We find that
the parameters converge to unique and physically plausible values.
}}
\label{fig:corner}
\end{center}
\end{figure*}

One-sigma credibility intervals
for these parameters in each of our ten model fits are presented in 
Table~\ref{tab:posteriors}, and an example posterior distribution, for the 
single-population Zipfian model, is presented in Figure~\ref{fig:corner}.

\begin{table*}
\caption{\emph{One-sigma credibility intervals of the model parameters for each of the ten multiplicity models. Recall that $\beta$ is defined differently for each model, as described in Section~\ref{sub:models}.}} 
\centering 
\begin{tabular}{l l l l l l l l}
\hline\hline 
Model & $\alpha_{\mathrm{small}}$ & $\alpha_{\mathrm{big}}$ & $R_{\mathrm{crit}}$\,[$R_{\oplus}$] & $\sigma_R$\,[$R_{\oplus}$] & $\beta$ & $\sigma_I$\,[$^{\circ}$] & $f$\\
\hline 
Constant & $0.22^{+0.42}_{-0.29}$ & $3.90^{+1.78}_{-1.49}$ & $2.69^{+0.61}_{-0.57}$ & $0.29^{+0.29}_{-0.20}$ & $4.70^{+0.54}_{-0.98}$ & $4.15^{+0.68}_{-0.72}$ & -\\
Uniform & $2.40^{+0.63}_{-0.49}$ & $2.61^{+1.58}_{-0.98}$ & $2.53^{+0.46}_{-0.37}$ & $0.25^{+0.20}_{-0.16}$ & $0.51^{+1.25}_{-1.77}$ & $5.80^{+0.46}_{-0.49}$ & -\\
Poisson & $-0.72^{+0.22}_{-0.13}$ & $4.52^{+3.43}_{-1.63}$ & $3.73^{+1.35}_{-1.00}$ & $1.20^{+0.27}_{-0.64}$ & $5.07^{+1.36}_{-1.35}$ & $0.02^{+0.05}_{-0.02}$ & -\\
Exponential & $0.53^{+0.66}_{-0.36}$ & $4.55^{+1.06}_{-0.79}$ & $2.49^{+0.31}_{-0.28}$ & $0.19^{+0.21}_{-0.13}$ & $1.65^{+0.44}_{-0.27}$ & $2.62^{+0.95}_{-0.90}$ & -\\
Zipfian & $0.60^{+0.60}_{-0.36}$ & $4.72^{+1.00}_{-0.78}$ & $2.53^{+0.27}_{-0.29}$ & $0.14^{+0.16}_{-0.09}$ & $0.86^{+0.28}_{-0.29}$ & $2.04^{+0.76}_{-0.69}$ & -\\
\hline
Di-Constant & $-0.17^{+0.23}_{-0.24}$ & $3.42^{+0.98}_{-0.75}$ & $2.83^{+0.33}_{-0.32}$ & $0.14^{+0.20}_{-0.10}$ & $4.90^{+0.40}_{-0.37}$ & $2.69^{+0.33}_{-0.55}$ & $0.72^{+0.04}_{-0.04}$\\
Di-Uniform & $1.88^{+0.58}_{-0.46}$ & $3.94^{+1.25}_{-0.95}$ & $2.55^{+0.35}_{-0.32}$ & $0.19^{+0.22}_{-0.14}$ & $1.98^{+0.64}_{-0.66}$ & $4.03^{+0.57}_{-0.51}$ & $0.65^{+0.06}_{-0.08}$\\
Di-Poisson & $0.14^{+0.29}_{-0.28}$ & $5.22^{+1.30}_{-0.75}$ & $2.61^{+0.33}_{-0.29}$ & $0.14^{+0.17}_{-0.09}$ & $5.16^{+1.21}_{-1.18}$ & $0.80^{+0.30}_{-0.33}$ & $0.55^{+0.04}_{-0.05}$\\
Di-Exponential & $0.65^{+0.69}_{-0.46}$ & $4.53^{+0.95}_{-0.69}$ & $2.49^{+0.30}_{-0.27}$ & $0.15^{+0.19}_{-0.11}$ & $1.51^{+0.40}_{-0.30}$ & $2.14^{+1.11}_{-0.97}$ & $0.44^{+0.11}_{-0.12}$\\
Di-Zipfian & $1.08^{+0.86}_{-0.54}$ & $4.47^{+1.14}_{-0.82}$ & $2.53^{+0.33}_{-0.29}$ & $0.17^{+0.18}_{-0.11}$ & $0.15^{+0.57}_{-0.82}$ & $2.87^{+0.95}_{-0.88}$ & $0.33^{+0.23}_{-0.22}$\\

\hline\hline 
\end{tabular}
\label{tab:posteriors} 
\end{table*}

\subsection{Model comparison}

To compare models, we use the Akaike Information Criterion, AIC
\citep{akaike:1974}. There are two major reasons behind this choice. First, the
AIC does not require that one of the models being tested is the correct model,
it merely asks which of the models is the closest approximation to the truth
(unlike the BIC; \citealt{schwarz:1978}). Second, the AIC does not functionally
depend on the sample size, which is somewhat ill-defined in our problem since our
inference employed a likelihood approximation. We therefore calculate the AIC
for each model by first finding the most probable realization from the 50,000
posterior samples, as defined by the distance function, and then using

\begin{align}
\mathrm{AIC} &= -2\log\hat{\rho} + 2k,
\end{align}

where $k$ is the number of free parameters used by each model. Since
we constructed our distance metric $\rho$ as a product of two likelihood-like
terms, $\rho$ approximates the likelihood in the AIC calculation above. For the 
non-mixture models, $k=6$ since we have free parameters $\beta$, $\sigma_I$,
$R_{\mathrm{crit}}$, $\sigma_R$, $\alpha_{\mathrm{small}}$ and $\alpha_{\mathrm{big}}$. The
dichotomous models add one extra free parameter, the fraction of single-planet systems $f$.

We assign uncertainties to our AIC scores through a bootstrapping procedure.
First, we split the chain up into $S$ segments. For each segment, we compute
the AIC, and estimate its standard deviation as 1.4826 multiplied by the median
absolute deviation. We repeat this procedure, varying $S$ from 2 to 50 in unity 
steps, and for the non-mixture models use the median score across all experiments. 
For the dichotomous models, we find that the scatter tends to decrease as we approach
small $S$, and thus we fit a simple quadratic model of scatter versus $S$ to
estimate the scatter at $S=1$. These uncertainty estimates, along with the
overall AIC scores, are compiled in Table~\ref{tab:AIC}.

\begin{table}
\caption{\emph{AIC scores and estimated uncertainties
for the ten different models used to describe the \kepler\
exoplanet multiplicity distribution.}} 
\centering 
\begin{tabular}{l l} 
\hline\hline 
Multiplicity Model & AIC \\ [0.5ex] 
\hline 
Constant	& $73.75 \pm 1.52$ \\
Uniform		& $64.17 \pm 0.82$ \\
Poisson		& $62.72 \pm 0.80$ \\
Exponential & $58.96 \pm 0.90$ \\
Zipfian		& $54.52 \pm 0.76$ \\
\hline
Di-Constant     & $55.92 \pm 0.20$ \\
Di-Uniform      & $55.02 \pm 0.25$ \\
Di-Poisson      & $55.87 \pm 0.15$ \\
Di-Exponential  & $55.39 \pm 0.57$ \\
Di-Zipfian	    & $54.55 \pm 0.97$ \\ [1ex]
\hline\hline 
\end{tabular}
\label{tab:AIC} 
\end{table}

Amongst the non-mixture models, Table~\ref{tab:AIC} shows that the Zipfian
distribution is preferred, favored over the next-best model
(exponential) with an odds ratio of $e^{(58.96-54.52)/2} = 9.21$. Since
all of these models have the same number of parameters, this preference
is purely driven by the much improved distance metrics. The constant model
is found to be the worst description of the multiplicity distribution,
disfavored versus a Zipfian model by a factor of 15,000.

Amongst the dichotomous models, the field is much more level, with all five
models roughly equally favorable to each other and also to
the single-population Zipfian distribution model. As a check on this, we also
tried computing the Savage-Dickey ratio by evaluating the posterior
density at $f_{\mathrm{single}}=0$. Only the di-exponential and di-Zipfian models had enough
samples around this region to reliably estimate the 
single-population model:dichotomous model odds ratio, yielding
ratios of 0.055 and 1.12, respectively. These are broadly
consistent with the AIC results of approximately equal weights
($=e^{-\Delta \mathrm{AIC}/2}$), demonstrating that the AIC approach is a suitable
approximation for this model selection problem.

On this basis, we conclude that the simpler hypothesis of a single population
model is not significantly rejected by the current data. A single Zipfian
distribution appears quite capable of describing the \kepler\ exoplanet
multiplicity distribution for FGK hosts.

\begin{figure*}
\begin{center}
\includegraphics[width=2.05\columnwidth,angle=0,clip=true]{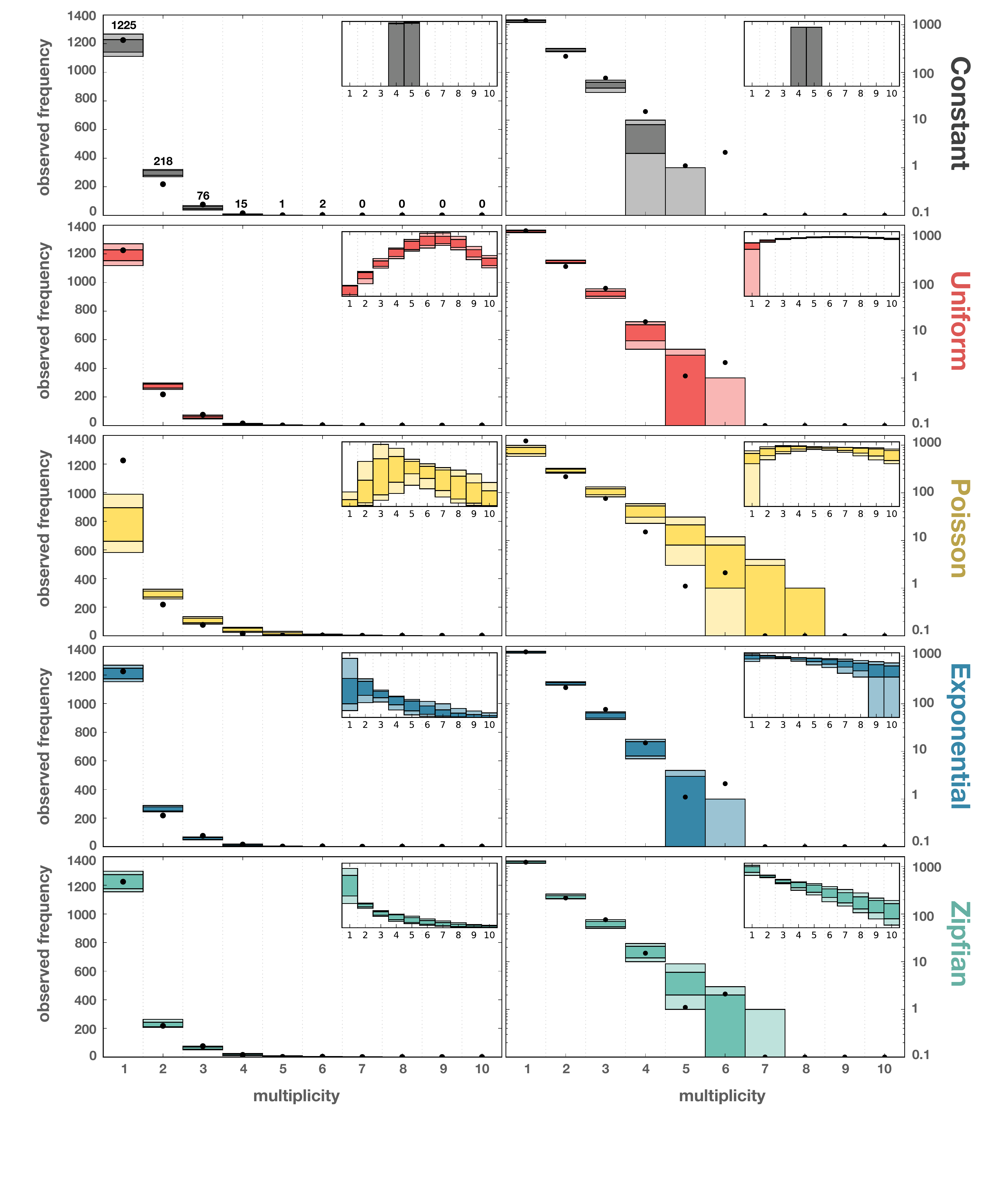}
\caption{\emph{
Left: Linear-scale histogram of the multiplicities of ``detected" simulated planetary
systems for \kepler\ FGK stars for the five single-population
models. We inset the underlying simulated multiplicity distribution
in each panel. The dark regions signify the 1-$\sigma$ credible
interval, and light regions give 2-$\sigma$. Black circles represent the
real observed \kepler\ sample. Right: Same as left except
log-scaled.
}}
\label{fig:results_singlepop}
\end{center}
\end{figure*}

\begin{figure*}
\begin{center}
\includegraphics[width=2.05\columnwidth,angle=0,clip=true]{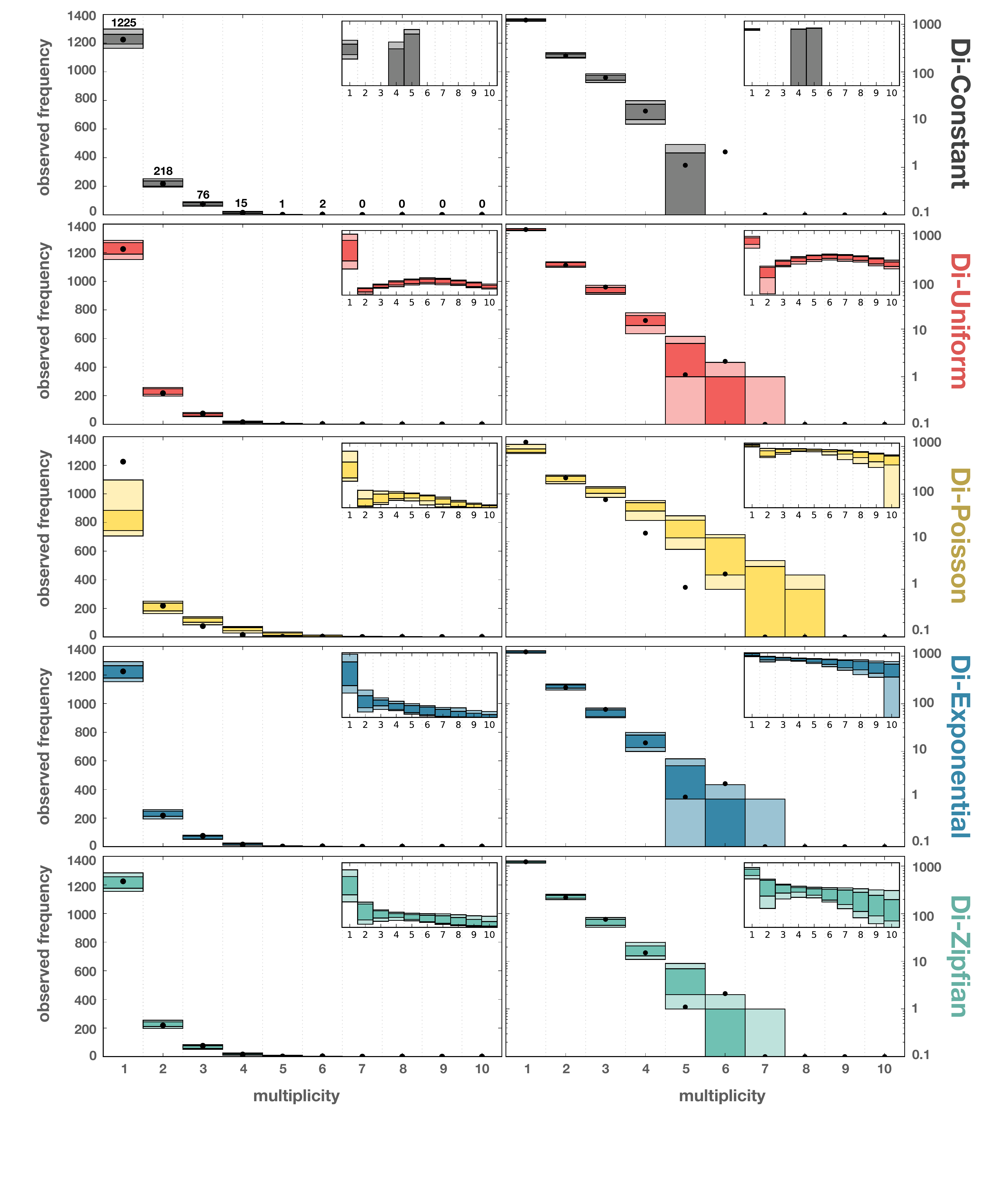}
\caption{\emph{
Left: Linear-scale histogram of the multiplicities of ``detected" simulated planetary
systems for \kepler\ FGK stars for the five dichotomous
models. We inset the underlying simulated multiplicity distribution
in each panel. The dark regions signify the 1-$\sigma$ credible
interval, and light regions give 2-$\sigma$. Black circles represent the
real observed \kepler\ sample. Right: Same as left except
log-scaled.
}}
\label{fig:results_dichotomous}
\end{center}
\end{figure*}

In Figures~\ref{fig:results_singlepop} and~\ref{fig:results_dichotomous}, we plot
the underlying simulated population of planetary systems (inset figures) and the ``detected"
subset of these systems, for all ten multiplicity models, 
to investigate the effect of detection biases on this subset
and to compare it to the real \kepler\ detections. Despite very different
underlying multiplicity models, the detected sample is qualitatively similar in all ten
cases: strongly peaked at $m = 1$, and falling off at higher multiplicities. It is not surprising
in this light that the Zipfian model performs best of the five single-population models,
as it has this general shape already, and that the dichotomous models (which by definition
include a peak at $m=1$) perform equally well.

Consequently, we conclude overall that the current \kepler\ data prefer multiplicity
models which peak at $m=1$, but have little distinguishing power between such models.
Choosing among them thus becomes a question of prior beliefs about the underlying
planet distribution---e.g., is there theoretical support for two planetary system
formation pathways?

In what follows, we investigate more fully the single-population
Zipfian multiplicity model, on the basis of its simplicity.

\subsection{Testing the inference framework}
\label{sub:fakes}

The use of ABC and also AIC model selection are both approximate tools
and thus one might reasonably question how robust they really are.
To test this, we decided to generate a total of 20 fake data sets where
the true multiplicity distribution is known and test how well we can recover
that true distribution using the same machinery used thus far.

We generate the first ten mock populations assuming a Poisson
multiplicity distribution. Every population has the same input parameters,
chosen to be close to the inferred solutions in our earlier fits, specifically
$\alpha_{\mathrm{small}} = 0.33$, $\alpha_{\mathrm{big}} = 5.0$,
$R_{\mathrm{crit}} = 2.5$\,$R_{\oplus}$, $\sigma_R = 0.05$\,$R_{\oplus}$, $\beta_{\mathrm{Poisson}} = 12.0$,
$\sigma_I = 2.0^{\circ}$. However, these ten fake ``observed'' data sets,
$\mathcal{D}_{\mathrm{obs}}'$, are slightly different to each other due to the
stochastic nature of the forward simulation. The second ten are generated in
the same way except we switch to a Zipfian distribution (replacing
$\beta_{\mathrm{Poisson}}$ with $\beta_{\mathrm{Zipf}}=1.0$).

We fit \textit{each} of these twenty data sets with two models: a Poisson and
a Zipfian. Thus, we should be able to test whether AIC scoring is able to pick
out the correct model in each case - a basic assumption upon which the previous
subsection rests. Second, we can test whether the inferred parameters (in cases
where the correct model is regressed) are compatible with the input values. In
this way, we can provide a detailed assessment of the validity of our inference
framework. To save computational time, we use 10,000 post-burn-in steps for
each MCMC fit.

\begin{table}
\caption{\emph{AIC scores for twenty fake data sets fitted using
two models. Boldened numbers indicate the favored model, which
equals the true model in 20/20 cases.}} 
\centering 
\begin{tabular}{l l l} 
\hline\hline 
Experiment & AIC (Poisson) & AIC (Zipfian) \\ [0.5ex] 
\hline 
Truth = Poisson \\
\hline

1   & \textbf{45.24}   & 56.57 \\
2   & \textbf{45.61}   & 53.93 \\
3   & \textbf{46.56}   & 56.25 \\
4   & \textbf{48.16}   & 61.57 \\
5   & \textbf{51.97}   & 63.72 \\
6   & \textbf{44.78}   & 58.66 \\
7   & \textbf{49.44}   & 54.01 \\
8   & \textbf{44.85}   & 57.28 \\
9   & \textbf{44.58}   & 53.27 \\
10  & \textbf{46.10}   & 52.38 \\

\hline
Truth = Zipfian \\
\hline

1   & 67.99   & \textbf{54.22} \\
2   & 96.51   & \textbf{58.96} \\
3   & 68.23   & \textbf{54.99} \\
4   & 91.57   & \textbf{64.30} \\
5   & 104.78  & \textbf{60.79} \\
6   & 61.67   & \textbf{54.25} \\
7   & 86.42   & \textbf{63.48} \\
8   & 72.13   & \textbf{59.85} \\
9   & 85.57   & \textbf{56.84} \\
10  & 91.90   & \textbf{58.16} \\ [1ex]
\hline\hline 
\end{tabular}
\label{tab:AICfakes} 
\end{table}

The AIC results, summarized in Table~\ref{tab:AICfakes}, show that the correct
model is identified in 20 out of 20 cases. In general, the
Zipfian model appears to be more flexible and gets closer to describing the
Poisson model than vice versa, likely as a result of the very harsh selection
functions which push the distributions towards an ostensibly monotonic form.
Nevertheless, the AIC scoring system appears to be a reliable tool for
identifying the best model.

\begin{figure*}
\begin{center}
\includegraphics[width=2.05\columnwidth,angle=0,clip=true]{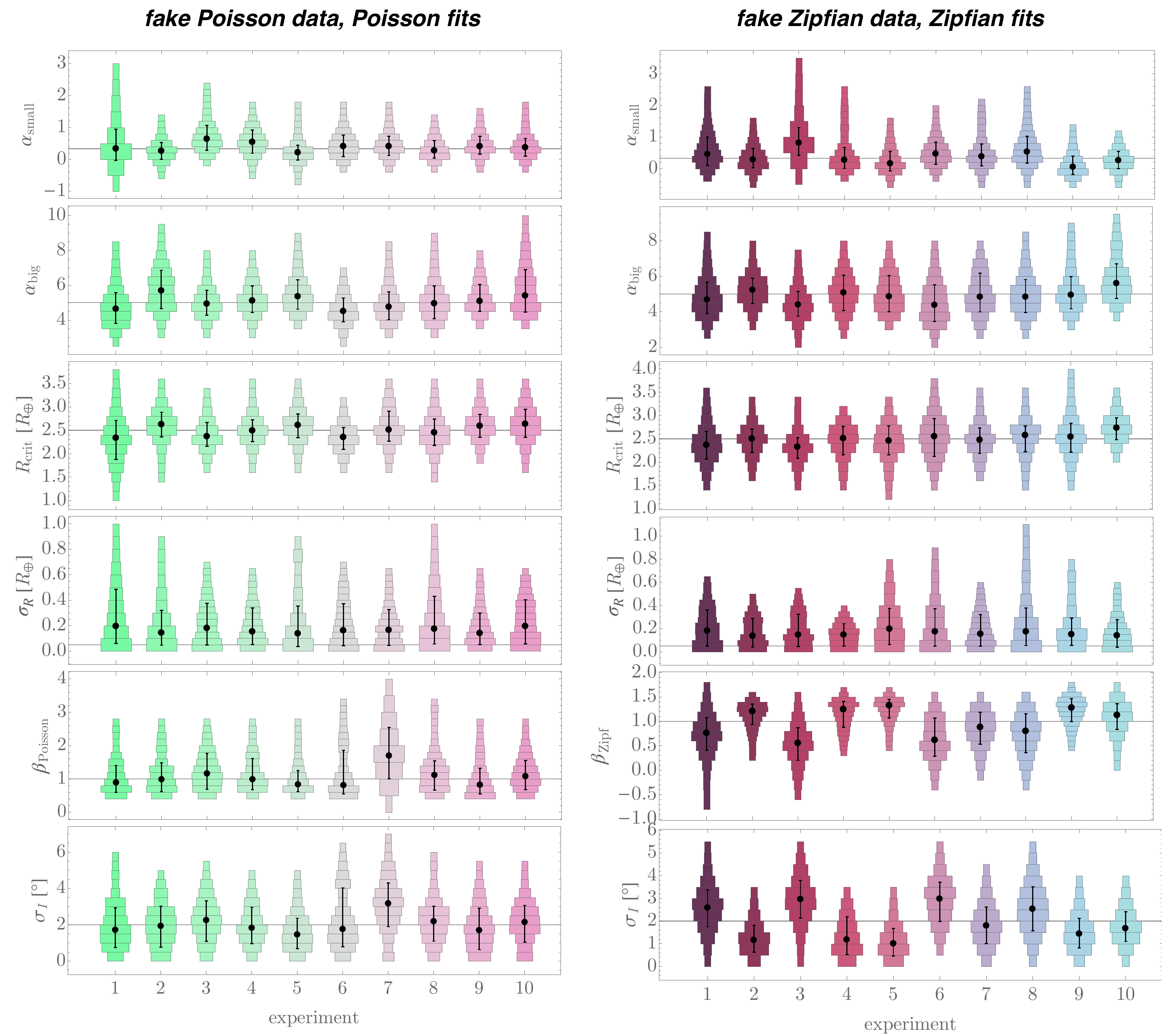}
\caption{\emph{
Violin plots comparing the retrieved a-posteriori distributions of the five
free parameters in our Poisson (left) and Zipfian (right) model, using the
described ABC inference framework. Each panel shows a different parameter, with
each labeled experiment representing an independently generated fake data set.
The injected truth is given by the horizontal lines, and the median and 1-sigma credibility band of the recovered parameter in each trial are plotted in black.}}
\label{fig:postfakes}
\end{center}
\end{figure*}

Comparing the actual parameters which result, as shown in
Figure~\ref{fig:postfakes}, we find good agreement between the results and the
injected truths: over the 10 fits to the 6 parameters, the recovered parameter is in $1\sigma$ agreement with the injected parameter in 60 of 60 cases for the Poisson trial and 54 of 60 cases for the Zipf trial. We also inspected the distances metrics versus parameter samples and verified
that the distance metrics approach their maximum around the true injected values,
as expected. This establishes that the ABC inference framework is able
to accurately recover the correct parameters, as well as being suitable for
model selection via the AIC.

\subsection{Properties of the preferred Zipfian distribution}

Given that the non-dichotomous Zipfian multiplicity model is the favored model
of this work, it is worthwhile to consider the parameters inferred from this
model. We show a corner plot of the joint posteriors in
Figure~\ref{fig:corner}, where the converged, unique nature of the inferred
solution is evident.

It is also instructive to compare the multiplicity of the generated systems
in our fits, versus the apparent multiplicity of these same systems after
being filtered through our mock \kepler\ pipeline. This is is shown in
Figures~\ref{fig:results_singlepop} and~\ref{fig:results_dichotomous}, 
where one can see how the true sample (inset figures) is considerably 
diminished as a result of detection bias.

\subsubsection{Interpreting the Zipfian slope}

When interpreting the inferred value of $\beta_{\mathrm{Zipf}}$, it
is worth highlighting that $\beta_{\mathrm{Zipf}}=-1$ leads to a precisely
uniform distribution, $\beta_{\mathrm{Zipf}}<-1$ leads to distributions
whose probability density monotonically increases with increasing multiplicity,
and vice versa $\beta_{\mathrm{Zipf}}>-1$ leads to distributions which decrease
with increasing multiplicity.

With a shape parameter of $\beta_{\mathrm{zipf}}=(0.86^{+0.28}_{-0.29})$,
our fit strongly favors a distribution which decreases with increasing multiplicity.
This is evident from Table~\ref{tab:zipfianvalues}, which presents the
relative frequency of each multiplicity as determined from the Zipfian
fit. Since Mercury is less than half an Earth radius and Mars is beyond
the period threshold used in this work, the Solar System is a 2-planet
system in our framework - a configuration found in $(20.2^{+1.3}_{-1.7})$\% of
our simulated FGK planetary systems. Packed, compact systems are rare, with 6-planets or
more constituting $\sim 13\%$ of our simulated FGK planetary systems.

\begin{table}
\caption{\emph{One-sigma credible intervals of the underlying planet multiplicity
for \kepler\ FGK stars with periods $6.25<(P/\mathrm{days})<400$ and sizes
$0.5<(R/R_{\oplus})<32$, as computed from the marginal posterior of
the favored non-dichotomous Zipfian model. Quoted scores are defined as the
percentage of FGK planetary systems with at least one planet in the quoted
period and radius range.}} 
\centering 
\begin{tabular}{l l} 
\hline\hline 
Underlying multiplicity & Credible interval \\ [0.5ex] 
\hline 
1 & $36.53_{-10.45}^{+10.08}$\% \\
2 & $20.23_{-1.68}^{+1.34}$\% \\
3 & $13.52_{-1.79}^{+1.40}$\% \\
4 & $9.59_{-2.20}^{+2.16}$\% \\
5 & $6.83_{-2.01}^{+2.30}$\% \\
6 & $4.80_{-1.65}^{+2.12}$\% \\
7 & $3.27_{-1.29}^{+1.77}$\% \\
8 & $2.17_{-0.94}^{+1.36}$\% \\
9 & $1.45_{-0.68}^{+1.02}$\% \\
10 & $1.02_{-0.51}^{+0.82}$\% \\ [1ex]
\hline\hline 
\end{tabular}
\label{tab:zipfianvalues} 
\end{table}

\subsubsection{Probability of additional planets in known systems}

One can also see that although the fraction of observed one-planet systems
represents 80\% of all systems (see Table~\ref{tab:kepcounts}), in reality
only 37\% of system are truly single (see Table~\ref{tab:zipfianvalues}).
Another way to think about this is that the Zipfian model finds that 37\% of
the simulated systems are genuinely single and all of these must yield
a planet with the correct geometry and detectability to have been ``detected''
by the simulated \kepler\ survey (else they would not have been included in the
final simulated catalog since our code would have not saved the realization).
Since 100\% of the 37\% truly single planet systems appear as singletons in
the final catalog, $80-37=43\%$ of the detected planets are multiple planet
systems for which only one planet was detected to transit. Thus, of the 80\%
of ostensibly single planet systems, 37/80=46\% are indeed genuinely single
and the other 43/80=54\% are yet-to-be-revealed multi-planet systems.

Accordingly, radial velocity follow-up of single transiting FGK \kepler\
systems has an a-priori $54.1_{-12.3}^{+12.9}$\% chance of detecting new planets
(after correctly propagating the uncertainties) with periods and radii in
our specified range. Given that there are 1225 single-planet systems in our
sample, that equates to $663_{-151}^{+158}$ hidden planets in the single-planet
systems.

\subsubsection{Total number of missing planets}

By calculating the total number of planets generated in the simulated systems,
we find that the Zipfian model predicts a total of $4359_{-717}^{+904}$ planets
residing around the 1537 FGK systems with known detections. Since only 1966
known planets reside around these stars, that means that there are
$2393_{-717}^{+904}$ hidden planets - which are expected to be dynamically
stable. Discovering these planets, perhaps through radial velocity follow-up, could increase the planet count around these stars by a factor of
$2.22_{-0.36}^{+0.46}$.

\section{Discussion}
\label{sec:discussion}

The principal finding of this work is that the observed multiplicities
of the \kepler\ FGK transiting systems can be well-explained without
invoking a dichotomous population model. Specifically, we find that
a Rayleigh mutual inclination distribution with a Zipfian multiplicity
distribution (the latter of which appears to have never been tried before)
is able to well-reproduce the observed catalog. This is not to say that
dichotomous models are disfavored---indeed, the single-population Zipfian
and the five dichotomous models perform equally well---only that invoking
a dichotomous population is not necessary to explain the detected
\kepler\ multiplicities. Furthermore, we find that the \kepler\ data do decisively
prefer multiplicity models peaked at $m=1$ over those peaked at higher
multiplicities.

\citet{bovaird:2017} also suggest that the dichotomous model may not be
necessary by considering an alternative inclination distribution. Since
inclination and multiplicity both affect the final catalog
\citep{tremaine:2012}, then it is certainly plausible then either (or
both) of these effects are able to explain the observed multiplicities
without invoking dichotomy. \citet{zink19} note, furthermore, that the
\kepler\ pipeline's decreased detection efficiency for multi-planet systems
could also explain the overabundance of \kepler\ singles. Finally, although 
our work centers on FGK stars, we highlight that \citet{gaidos:2016} also 
find that a dichotomous distribution may not be necessary by changing 
the underlying models in the case of M-dwarfs.

It is curious that Zipf's Law \citep{zipf:1935}, most commonly associated with
linguistics, works well for exoplanet multiplicities. Zipfian laws are
argued by \citet{aitchison:2016} to be natural outcomes of systems involving
a large number of latent variables, and this may represent another example.
Extending our analysis to M-dwarfs, particularly from TESS, will provide a good
test as to whether the Zipfian model can persist in the face of new data.

Using our preferred model, we are able to make predictions about the numbers
of missing planets. For example, we predict that 7 or more planet systems
are rare, with just 7.9\% of detected systems being so packed. This is in sharp
contrast to \citet{mulders:2018}, who recently estimated that 42\% of Sun-like
stars have nearly coplanar planetary systems with 7 or more exoplanets.
Although our numbers are not measuring precisely the same quantity, it would
be difficult to reconcile the \citet{mulders:2018} value with our estimates
given the stark paucity of such systems in our observed sample.

Our model does predict a large number of missing planets, $\simeq 2400$
around the 1537 host stars considered, of which some $\simeq 660$ reside in
ostensibly single-transiting-planet systems. It may therefore be possible to
test the predictions of these models by conducting radial velocity follow-up of
the \kepler\ field in the future to measure the true multiplicities.
	
\section*{Acknowledgments}

ES, DK, \& MC acknowledge support from the Columbia University Data Science
Institute ``Seed Funds Program''. Thanks to members of the Cool Worlds Lab
for useful discussions in preparing this manuscript. We are grateful to
the anonymous reviewer for their constructive feedback, and to Jessi
Cisewski-Kehe for guidance on ABC.

%

\bsp
\label{lastpage}
\end{document}